# Using the SP!CE Framework to Code Influence Campaign Activity on Social Media: Case Study on the 2022 Brazilian Presidential Election

Alexander Gocso[†], Claudia Perez Brito[†], Bryan Ruesca[†], Allen Mendes[†], & Mark A. Finlayson[‡]

[†]Jack Gordon Institute for Public Policy
[‡]Knight Foundation School of Computing and Information Sciences
Florida International University
Miami, FL 33199
`markaf@fiu.edu`

December 12, 2023

## Abstract
We describe a case study in the use of the Structured Process for Information Campaign Enhancement (SP!CE, version 2.1) to evaluate influence campaigns present in the 2[nd] round of the Brazilian presidential election in 2022 October. SP!CE is a US-military focused framework for describing both friendly and adversary actions in influence campaigns, and is inter-operable with the Disinformation Analysis and Risk Management (DISARM) framework. The purpose of the case study is to demonstrate how SP!CE can be used to describe influence campaign behaviors. We selected the Brazilian election as the target of the case study as it is known that there were significant amounts of mis- and disinformation present on social media during the campaigns. Our goal was to demonstrate how SP!CE could be applied in such a context, showing how social media content could be aligned with information campaign behaviors and how such an alignment can be used to analyze which mis- and disinformation narratives were in play. Additionally, we aim to provide insights on best practices regarding how to apply the framework in further research. We release the coding and screenshots of the relevant social media posts to support future research.[1]

## Introduction
Modern digital platforms—in particular, social media platforms—have enabled the broad-based global dissemination of information, but unfortunately have also enabled the spread misinformation and disinformation, which have significant malign effects on society. Here we define *misinformation* as misleading information without the intent to cause harm, and *disinformation* as deliberately created false or misleading information.[2] Mis- and disinformation can influence people to act in ways contrary to their own or society's interests. For example, during the COVID-19 pandemic mis- and disinformation led people to refuse vaccines and resist wearing masks.[3]

---

[1] Data can be downloaded from https://doi.org/10.34703/gzx1-9v95/8PC8JY.
[2] What is Misinformation/Disinformation?, Purdue University, https://www.lib.purdue.edu/misinformation-training/training-module/what-is-misinformation. Last accessed June 2, 2023.
[3] Ferreira Caceres, M. M., Sosa, J. P., Lawrence, J. A., Sestacovschi, C., Tidd-Johnson, A., Rasool, M. H. U., Gadamidi, V. K., Ozair, S., Pandav, K., Cuevas-Lou, C., Parrish, M., Rodriguez, I., & Fernandez, J. P. (2022). The impact of misinformation on the COVID-19 pandemic. *AIMS public health*, *9*(2), 262–277. https://doi.org/10.3934/publichealth.2022018. Last accessed May 28, 2023.

Notably, when it comes to mis- and disinformation, prevalence should not be conflated with impact[4]. Despite the lack of unequivocal evidence that large amounts of mis- and disinformation result in effective influence, a variety of actors continue to conduct influence campaigns spreading mis- and disinformation in an effort to encourage outcomes favorable to their mission and goals. For example, the U.S. Office of the Director of National Intelligence concluded Russian President Putin authorized an influence campaign aimed at the 2020 U.S. presidential elections with the goal of "denigrating President Biden's candidacy and the Democratic Party, supporting former President Trump, undermining public confidence in the electoral process, and exacerbating sociopolitical divisions in the US.", in the belief that these effects were beneficial to Russia's strategic goals.[5]

As adversaries continue to use and refine strategies in the execution of influence campaigns to spread disinformation that aligns with their geopolitical goals, it is critical for governments and organizations that value democratic principles and the dissemination of accurate information to invest resources in identifying mis- and disinformation behaviors to identify and mitigate any potential impacts. One challenge in this process is establishing a uniform vocabulary to describe mis- and disinformation behaviors. The SP!CE framework, developed jointly by MITRE and FIU, aims to fill part of that gap. SP!CE is inspired by the MITRE ATT&CK framework for describing observables in the cybersecurity domain, builds upon the prior AMITT framework for describing online influence behaviors, and is aligned with the DISARM framework, which is an open-source framework for describing observed influence campaign behaviors.[6]

The SP!CE 2.1 framework, used for this case study, is comprised of four phases—Plan, Enable, Engage, Assess—which correspond to natural stages in an influence campaign. Within each phase there are different tactics, and within the tactics there are corresponding techniques. In total the framework contains 23 tactics and 106 techniques. The purpose of the case study presented here is to provide an example of how to apply SP!CE to real-world data, providing additional labeled data to support the building of SP!CE-related products, and also show how such an analysis can support identification of adversary narratives in mis- and disinformation campaigns.

We chose the 2nd round of Brazilian election of 2022 as the target for our case study because it is known that were significant amounts of mis- and disinformation present during the campaign[7]. In the Spring of 2023, we collected 450 social media artifacts containing apparent mis- and disinformation from social media platforms and channels that were posted during a month-long period between September 2 and October 2, 2022 (431 tweets, 13 news articles from Russia Today, and 6 other online news articles). We then exhaustively classified the artifacts according to which SP!CE tactics and techniques we inferred were in use. Our analysis revealed 450 artifacts reflecting tactics in the SP!CE Engage phase and 73 in the Enable phase. We did not discover any activities which could be marked as a part of the Plan or Assess phase, which is consistent with our expectations that these phases rarely result in observable social media activity.

---

[4] Altay, S., Berriche, M., & Acerbi, A. (2023). Misinformation on Misinformation: Conceptual and Methodological Challenges. *Social Media + Society*, *9*(1). https://doi.org/10.1177/20563051221150412. Last accessed May 28, 2023.

[5] Foreign Threats to the 2020 US Federal Election, https://www.dni.gov/files/ODNI/documents/assessments/ICA-declass-16MAR21.pdf. Last accessed June 2, 2023.

[6] DISARM Foundation, DISARM Framework Description, https://www.disarm.foundation/framework. Last accessed May 29, 2023.

[7] Social media platforms crack down on fake news ahead of Brazil election, Financial Times, https://www.ft.com/content/afd5d2dc-af51-411e-9eb4-eed4b28db55c. Last accessed June 2, 2023.

## Methods

Brazil is the fourth-largest social media market in the world.[8] In 2022 there were approximately 19.05 million Twitter users in Brazil, and it is the eighth largest social media platform in the country after WhatsApp, Instagram, Facebook, TikTok, Facebook Messenger, Telegram, and Kuaishou[9]. We selected Twitter for this study given its accessibility when searching and filtering content. Additionally, Twitter is not as quick to remove inauthentic content as other platforms, giving us additional time to collect relevant artifacts. We also added a small handful of news articles that were relevant to the tweets we found.

We first used search on Twitter and Google to find content relevant to the election. For each tweet or news article found, we evaluated it as to whether it contained known mis- or disinformation themes, or seemed to implement a tactic or technique found in the SP!CE framework. As noted, this search-and-filter approach resulted in 450 artifacts (431 tweets and 19 news articles); the size of the set was limited by project time and budget considerations.

We primarily made use of Twitter's advanced search option to filter content. Specifically, during the Spring semester of 2023 (January through April) we searched for individual terms or hashtags with specific date constraints (September 1 to October 1, 2022, inclusive). It's important to note that using Twitter's advanced search option does not return all tweets that match the search parameters. Instead, Twitter filters search results for "quality Tweets and accounts". This means Twitter tries to filter out potentially automatic, inauthentic, and/or duplicate tweets.[10]

### Search Terms & Hashtags

Table 1 lists search terms and hashtags we used to filter content through Twitter's advanced search function.

| Keywords | Hashtags | |
|---|---|---|
| Bolsonaro        | `#brazil/ #brasil`         | `#brasil`            |
| Lula             | `#brazilnasruas`           | `#eleicoes2022`      |
| Brazil           | `#patriotabrasiljmb`       | `#lulacriminoso`     |
| Brazil elections | `#brazilelections`         | `#outlula`           |
| Brasil           | `#brazilianstaketostreets` | `#lulano1oturno`     |
| Brasil elecciones| `#sosffaa`                 | `#bolsonarogenocida` |
| Brasil eleicao   | `#brasildecide`            | `#bolsonaro`         |
|                  | `#brasilelige`             | `#lula`              |

Table 1: List of Keywords and Hashtags used in the search. Hashtags were only used in the Twitter search.

---

[8] Leading countries based on number of Twitter users as of January 2022, Statista, https://www.statista.com/statistics/242606/number-of-active-twitter-users-in-selected-countries/. Accessed May 15, 2023.
[9] Most popular social media platforms in Brazil as of 3rd quarter 2022, Statista. https://www.statista.com/statistics/1307747/social-networks-penetration-brazil/. Accessed May 15, 2023.
[10] About search rules and restrictions, Twitter, https://help.twitter.com/en/rules-and-policies/twitter-search-policies. Accessed May 15, 2023.

The keywords, but not the hashtags, were used in Google search to identify our small set of relevant news articles. They keywords are relatively simple and capture both English, Spanish, and Brazilian Portuguese terms for *election*, as well as the last names of the two candidates. The hashtags were a selection of the most well-known 2022 election related hashtags known to us at the start of the project.

**Cataloguing Content**

For each combination of a keyword or hashtag with our date range, we examined the results and attempted to determine if the content reflected a tactic and/or technique in the SP!CE framework. For some tactics and techniques, historical characteristics of the posting are relevant (i.e., how likely is it to be a bot account); we evaluated this using TruthNest[11], a Twitter site for analyzing inorganic activity, suspicious content, and bot indicators. If in our judgement an artifact did reflect a tactic and/or technique, we captured the following information:

- Account Name
- Website (if a news article)
- Content URL link
- Posting date
- Full text in original language (if applicable)
- A short summary of the post in our own words
- Screenshot
- Identified tactic
- Identified technique
- Identified sub technique (if applicable)

# Findings

**1) Enable Phase**

**1.1) Establish Information Assets and Intermediaries Tactic**

**1.1.1) Create Online Entities Technique**

The "Create Online Entities" technique encapsulates the creation of online digital resources that will enable automatic or semi-automatic broadcast of messages during the Enable phase. To determine if accounts were bots, cyborgs, anonymous accounts, we first used analysis tools such as TruthNest, BotSentinel[12], Botometer[13], FollowerAudit[14], and Heavy.ai[15] to generate independent verification of these entities' inorganic activity. By supplying the Twitter account's user handle, these tools aggregate the account's content and provide a variety of statistics, such as the users tagged most often and the most frequently used hashtags (Figure 1). For example, TruthNest provides a percentage estimate on how likely an account is a bot (Figure 2). The site does this by aggregating activity such as posts, frequency of posts, tagged accounts, and more.

---

[11] TruthNest, https://app.truthnest.com/ Last accessed July 29, 2023
[12] Botsentinel, https://botsentinel.com/, Last accessed July 29, 2023
[13] Botometer, https://botometer.osome.iu.edu/, Last accessed July 29, 2023, Website no longer available
[14] FollowerAudit, https://www.followeraudit.com/, Last accessed July 29, 2023
[15] Heavy.ai, https://www.heavy.ai/, Last accessed July 29, 2023

Each tool has limits to their capabilities which is why we used several in combination. For example, FollowerAudit only lets you audit Twitter accounts that have less than 5,000 followers if you're using the free version of the tool.

If an account showed signs of bot-like behavior as assessed by these tools, we examined the account creation date: if the account was created in 2022 before the election (January to October 2022), we hypothesized that these accounts were created for the purpose of carrying out influence campaigns for the 2022 election.

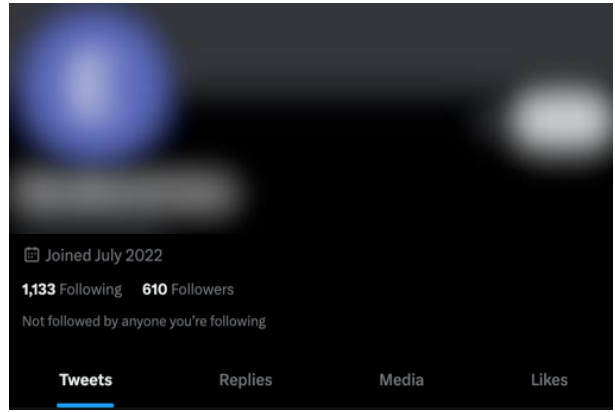

Figure 1: Twitter profile of an account likely engaging in inorganic activity relating to the Brazilian election (2022).

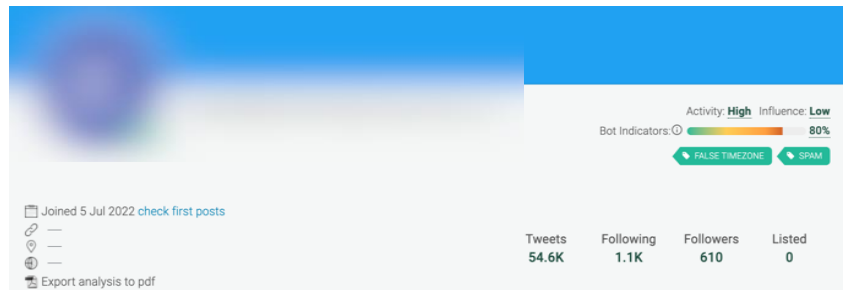

Figure 2: Partial view of a Truthnest analysis result showcasing the tool's percentage bot indicator. This image indicates the analysis on the Twitter account in Figure 1.

## 1.2) Establish Legitimacy Tactic

### 1.2.1) Create Localized Content Technique

The "Create Localized Content" technique involves developing content that appeals to specific communities by varying the language or message. This includes targeting specific age groups by developing content specific to social media platforms those groups might use most frequently. Our evidence showcased the use of this technique through the publication of a TikTok video in Brazilian Portuguese. Based on the post's language it is reasonable to infer that it was meant for a Brazilian audience. Additionally, given over 50 percent of TikTok users are below the age of thirty-four[16] this post was likely intended to target younger audiences. The video shows a crowd and claims they are all Bolsonaro supporters (Figure 3). Upon further research, we found the video is originally from a Brazil Independence Day celebration but it was used throughout the election cycle erroneously attributed to different sides[17]. In this case, the intent is to convey that support for Bolsonaro is large scale. In addition, his supporters are trying to use the video to aid their claims about the prospect of a fraudulent election should he lose.

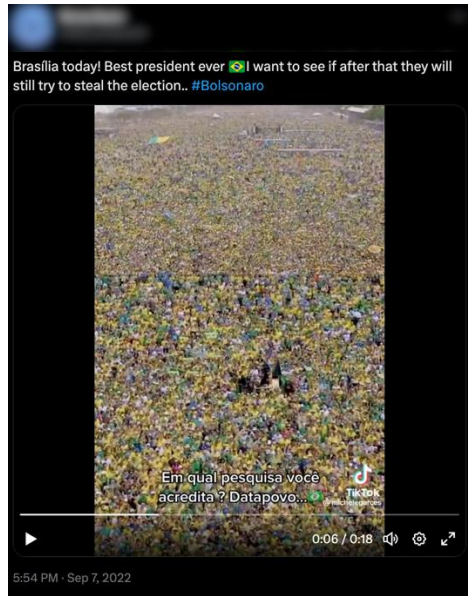

Figure 3: Post sharing a TikTok video with caption "Best president ever. I want to see if after that they will try to steal the election"

### 1.2.2) Leverage Existing Biases Technique

We hypothesized that we observed the "Leverage Existing Biases" technique with content that targeted information consumers who likely already had existing beliefs on certain issues. In one post, we observed the use of patriotism for Brazil and disagreement with European Union interests as a motivation to vote for Bolsonaro (Figure 4). This post claimed Lula would "kill" the "God-fearing and family-centered" Brazilian identity, and he would "sell out to Macron/European Union interests." By using this social framing, the narrative aimed to denigrate EU-aligned interests to further paint them as something that could undermine the Brazilian identity. It also allegedly claimed Brazilian values are at risk and thus implied the way to save the country would be to vote for Bolsonaro.

---

[16] TikTok Users by Age, https://www.oberlo.com/statistics/tiktok-age-demographics. Last accessed July 4, 2023.
[17] Fact check: Video shows Brazil Independence Day celebration, not post-election demonstration, USA Today, https://www.usatoday.com/story/news/factcheck/2022/12/15/fact-check-video-shows-brazil-holiday-celebration-not-demonstration/10885050002/, Last accessed June 3, 2023

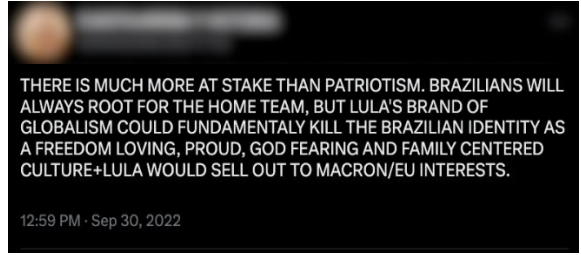

Figure 4: Post using the "Leverage Existing Bias" by focusing on patriotism for Brazil through conservative values.

### 1.3) Emplace Sensors Tactic

### 1.3.1) Use Poll/Survey Data Technique

For this technique, we discovered the frequent use of unofficial polls and surveys on Twitter. These posts attempted to garner user engagement and sway public perception based on misleading results. Despite being unrepresentative in their sampling, the poll results were presented as fair and representative of the Brazilian population. In addition to being Enable phase tactics, posts like these can have an Enable phase characteristic in that they work to dissuade users who prefer a particular candidate if their chosen candidate appears to be losing. Similarly, it can work to motivate undecided voters to choose the popular candidate. They can function as echo chambers and, depending on which accounts vote, they can also be easily manipulated. For instance, these inaccuracies such as those highlighted in tweets from one user used a poll to showcase support divided among candidates. The result (out of 1, 813 votes) in this post showed Bolsonaro with 86.4% of the vote. The malign actors can be grouped by various anonymous accounts that have bot like activity (Figure 5). Although we noted this Enable phase function, there was no corresponding tactic under which to place such activity in the SP!CE Enable phase.

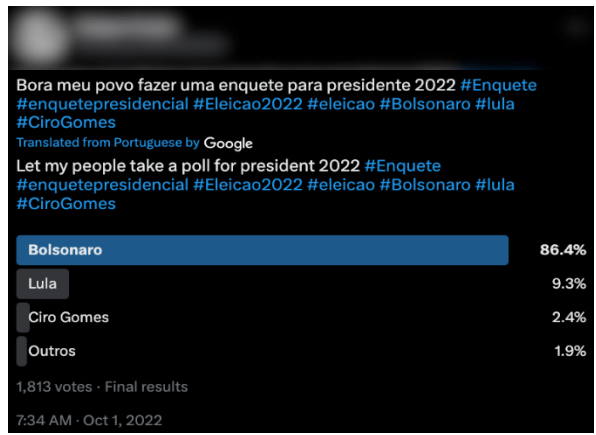

Figure 5: Post using an unofficial poll.

### 2) Engage Phase

### 2.1) Distort Existing Narratives Tactic

### 2.1.1) Amplify Conspiracy Theories Technique

The Brazilian presidential election has been the subject of many conspiracy theories, with the most prominent being that the electronic voting machines used to record votes were rigged and inefficient. These claims were mostly perpetuated by accounts that shared content claiming the election would be stolen from Bolsonaro and polls were pro-Lula. These claims create a perception the Brazilian electoral system favors Lula with manipulated voting machines. Messages in these posts also aimed to justify potential demonstrations should the outcome not be favorable to their preferred candidate. For example, one post claimed it would be time for Bolsonaro supporter to take to the streets should Lula win (Figure 6), while another post said those who consumer certain news outlets make Bolsonaro appear as if he is again democracy (Figure 7). It is interesting to note that the post shown in Figure 6 might also be classed under the "Call to Action" Tactic in the Engage phase.

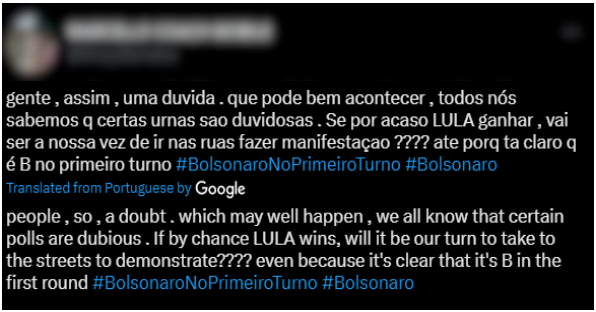

Figure 6: Post claims some polls are uncertain and alluding to potential protests should Lula win the election

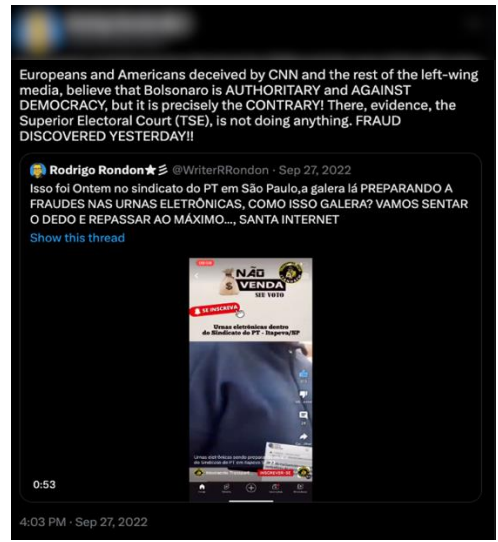

Figure 7: Post claiming those consuming outlets like CNN believe Bolsonaro is authoritarian and that election fraud was discovered.

## 2.1.2) Post Provocative Content Technique

Content that we classified as effecting this technique were mostly caricatures, photoshops, and out-of-context information that seemed to be deliberately provocative to drive engagement. Some accounts engaged in this technique would produce edited images all carrying the same anti-Bolsonaro and pro-Lula narrative. This content seemed to likely divide people and increase engagement. Their success is characterized by their ability to create "clickbait" content. For example, one post published a short video of a naked women being lashed for her alleged candidate preference (Figure 8). Other posts were memes ridiculing Bolsonaro as a crying baby (Figure 9) or as Harry Potter villain Voldemort (Figure 10).

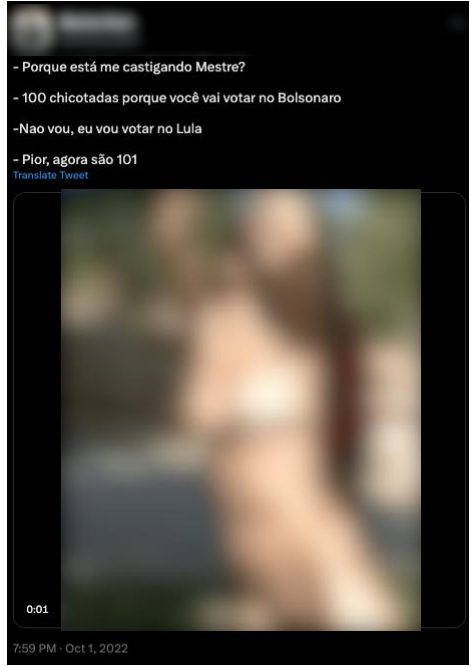

Figure 8: Post showcasing a man whipping a woman with the caption "Why are you punishing me master? 100 lashes because you are going to vote for Bolsonaro. No, I'm going to vote for Lula, Worse, now there are 101 [*sic*].

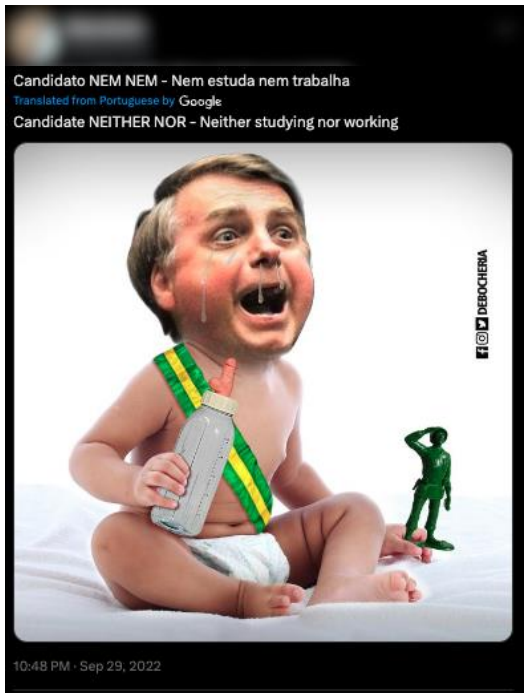

Figure 9: Meme of Bolsonaro as a crying baby.

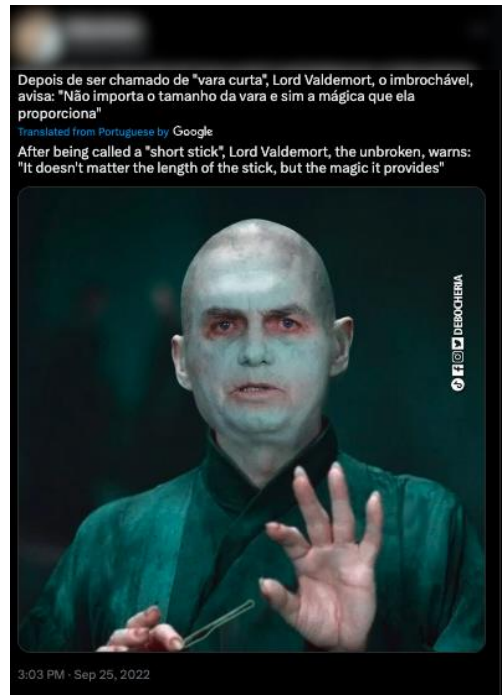

Figure 10: Meme of Bolsonaro as the Harry Potter villain Voldemort.

### 2.1.3) Reframe Context Technique

Reframing context refers to removing an event from its surrounding context to distort its intended meaning. Rather than deny that an event occurred, reframing context frames an event in a manner that may lead the target audience to draw a different conclusion about its intentions. Artifacts we classified as using this technique centered on misattributing the actions of celebrities by conveying, falsely, that they support for a candidate. One post shared more than once was an image of American basketball player Lebron James with his hand as an L shape incorrectly claiming he was supporting Lula when the image was taken from elsewhere (Figure 11 & Figure 12). Another example of the technique was when users tied American singer, Taylor Swift, to Lula. Images would use Swift's promotional content from her *Red* album and falsely convey she was supporting Lula (Figure 13). One post used an image of Taylor Swift with the number 13, her favorite number, to also tie she falsely was supporting Lula as that was his candidate number (Figure 14). It also used American actress Selena Gomez in a similar fashion.

Another example of this technique involved placing a photo of Bolsonaro in military uniform side-by-side with Lula's arrest mugshot (Figure 15 & Figure 16). The dichotomy between the two images aimed to further a pro-Bolsonaro narrative showcasing him in an honorable light as a public servant while painting Lula as a criminal. Depending on who they support, users created visual content to further the divide and create a distinctive outlook for both candidates.

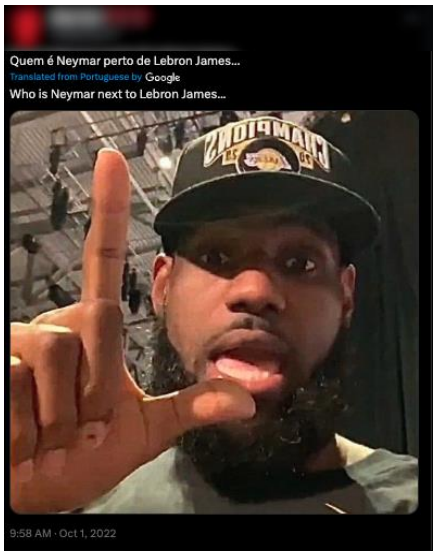

Figure 11: Post of American basketball player, Lebron James, with an L-shaped hand sign erroneously referring he is supporting Lula.

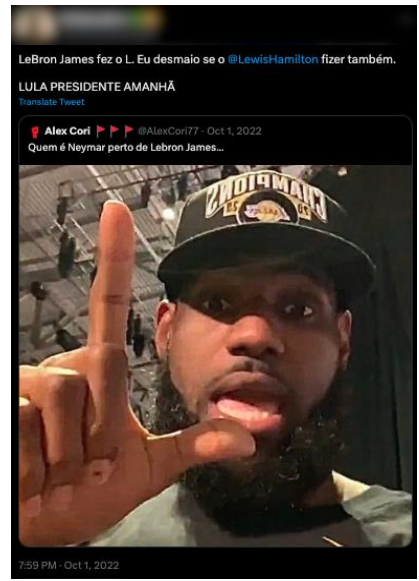

Figure 12: Another post showing the same image as in the previous figure.

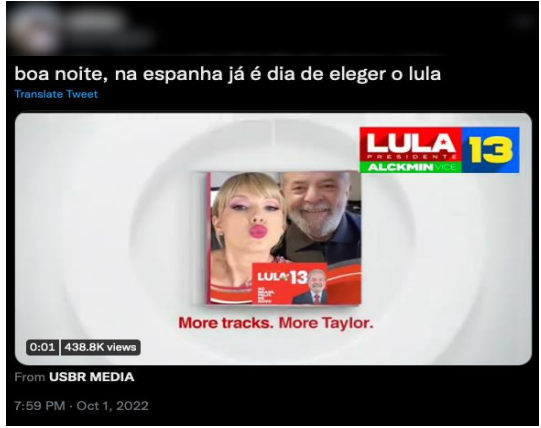

Figure 13: Post with an edited image of American singer, Taylor Swift, with Lula erroneously conveying she supports him.

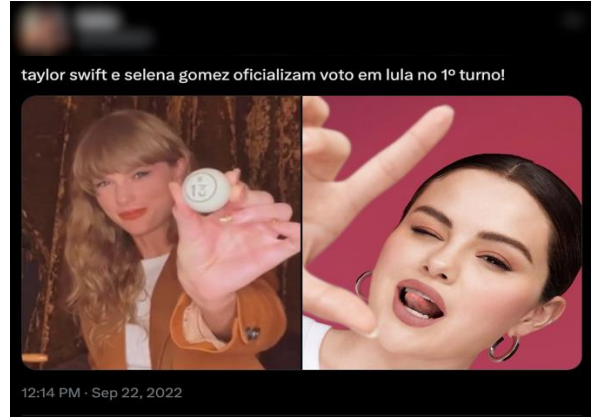

Figure 14: Post with Taylor Swift holding an item with the number 13, her favorite number, falsely conveying she supported Lula as 13 was his candidate number. The post also used American actress Selena Gomez in a similar manner.

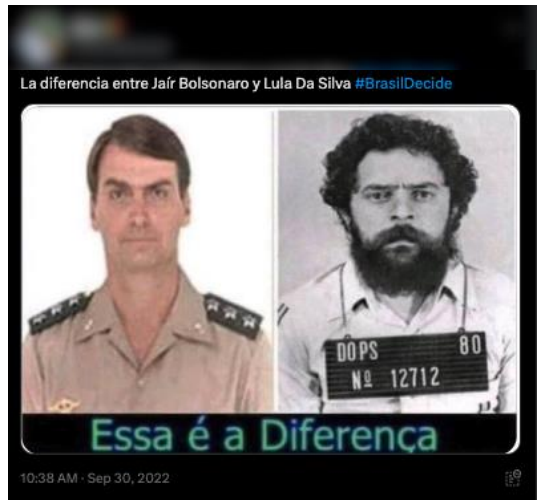

Figure 15: Post showcasing Bolsonaro's military headshot and Lula's mug shot with the intent to showcase the dichotomy between their characters and history. Subtitle translation: This is the difference.

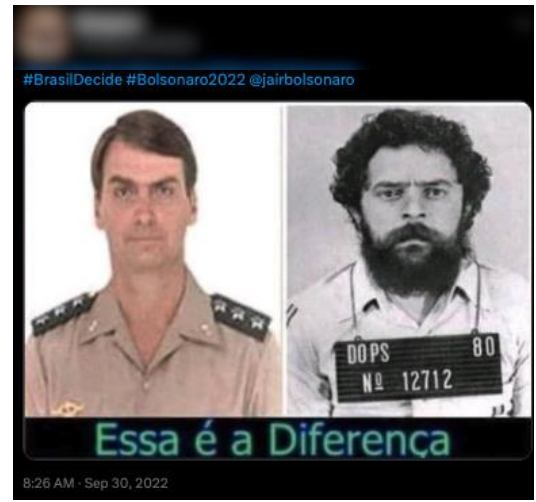

Figure 16: Another post showcasing the same differences between Bolsonaro and Lula.

## 2.2) Amplify Supporting Information (Maximize Exposure) Tactic

### 2.2.1) Exploit Platform-Specific Features Technique

Sharing memes was common for disseminating narratives of support and/or criticism for candidates and incentivizing sharing on Twitter. Most posts in this category demonized and ridiculed Bolsonaro, the right-wing candidate, using disingenuous and misleading memes and other images to nullify or discredit him. Notably, there were also posts that fit this technique that were also present in other techniques such as posting provocative content. Examples of posts that fit both include the photoshopped crying baby Bolsonaro image (Figure 17) and the Voldemort Bolsonaro image (Figure 18). Other posts with memes alluded to *Bolsonarismo*, the phenomenon tied to Bolsonaro's rise, being the same as the Nazi party (Figure 19 & Figure 20). Another post mocked Bolsonaro by editing him being thrown in a garbage truck (Figure 21). These memes work to share narratives in an exaggerated fashion.

In this technique, we observed accounts using disingenuous/misleading memes and other images to nullify or discredit candidates. Sometimes these visuals work to paint an exaggerated picture of the candidate and in turn ridicule them and vice versa. Individuals who agree with the message and tone of the visual can use this as steam to further disseminate the content. One post showed an edited image of Hitler on TIME magazine with the face of Bolsonaro superimposed in attempts to further vilify the presidential candidate. We can group these users' by identifying them as a conglomerate of smaller "personal" accounts who are ideologically aligned with one political candidate. This technique also featured the use of various hashtags to ensure the posts are aggregated with other content sharing the same hashtag (Figure 22). Some hashtags examples include #Boslonaro2022 and #Lulanuncamais which translates to 'Lula never again'.

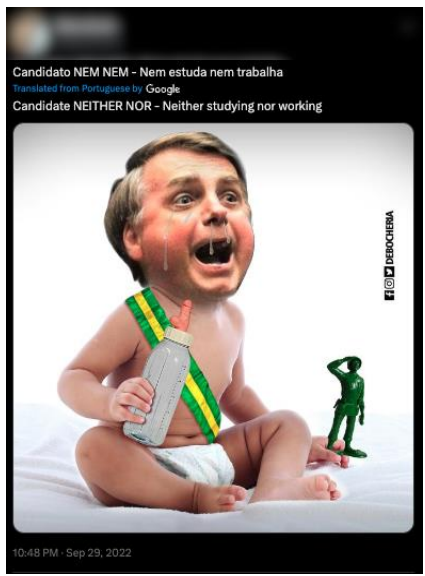
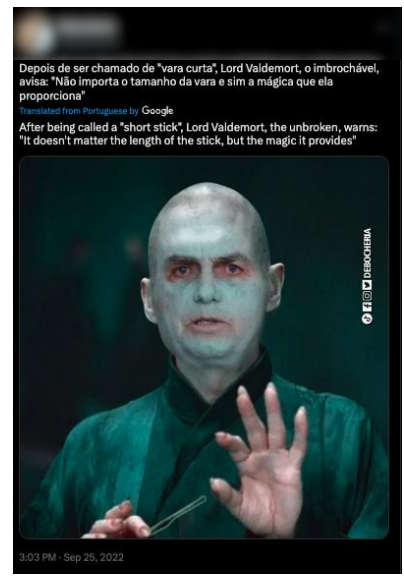

Figure 17: Meme of Bolsonaro as a crying baby.　　　Figure 18: Meme of Bolsonaro as the Harry Potter villain Voldemort.

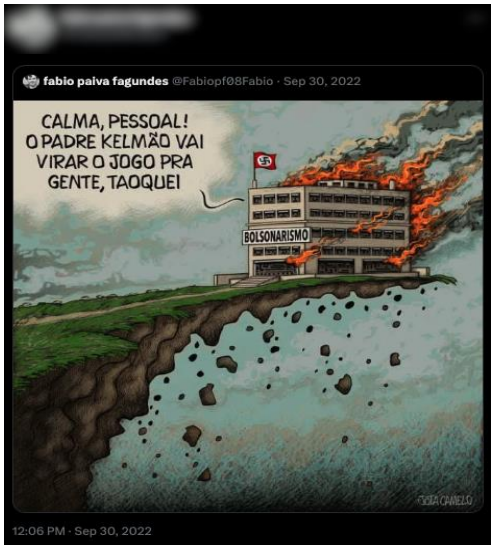

Figure 19: Meme of Bolsonarismo being a burning building at the edge of a cliff with a Nazi flag on the top.

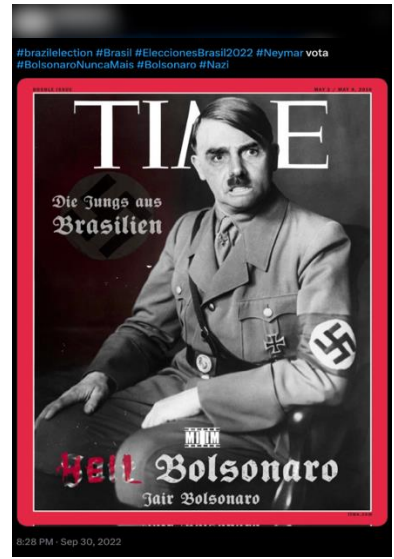

Figure 20: Meme of Bolsonaro's face superimposed onto Adolf Hitler's body appearing on the cover of a TIME magazine.

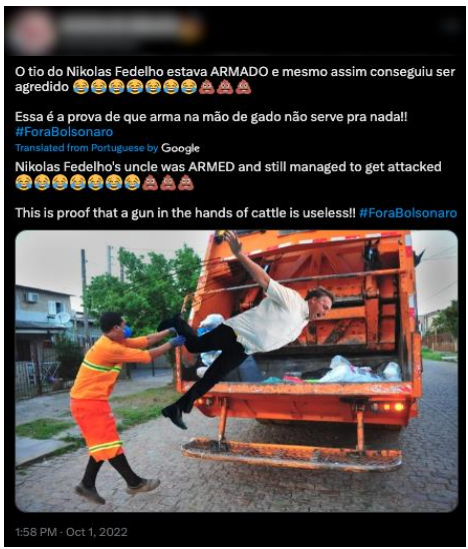

Figure 21: Meme of Bolsonaro being thrown into a dumpster truck.

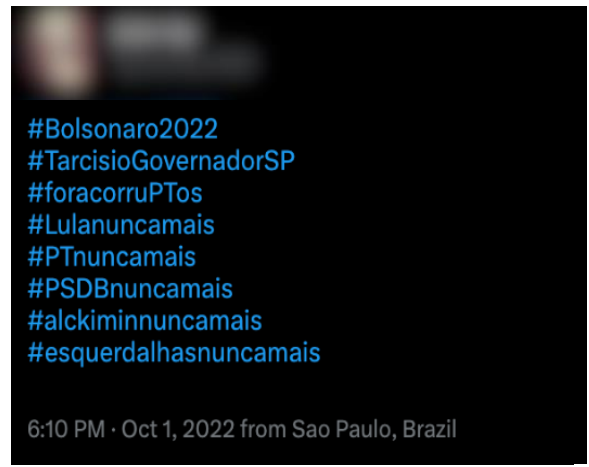

Figure 22: Post using hashtag spamming to make content more visible by riding popular hashtags.

### 2.2.2) Incentivize Sharing Technique

With this technique, users are encouraged to share content themselves by being permitted to tailor a message. The message overall helps disseminate the content that aligns with information operation goals. Posts in this technique we observed included the use of encouraging users to promote Lula with a customized profile picture ribbon through Twibbon, a brand overlay tool that is marketed for online campaigns (Figure 23 & Figure 24). This helped promote Lula as the preferred candidate.

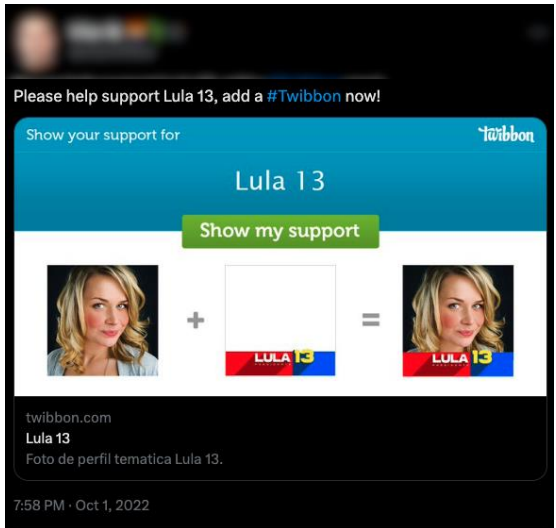
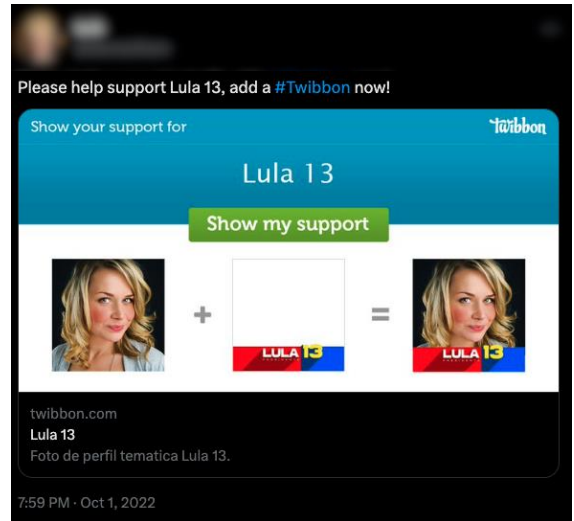

Figure 23: Post with link to Twibbon, a brand overlay tool, to showcase support for Lula.

Figure 24: Another post showcasing the same link to Twibbon.

### 2.2.3) Astroturfing Technique

The strategy aims to increase online exposure by incorporating additional supporting causes into campaigns. This can be achieved by disguising content to appear as if it belongs to a grassroots movement or organization whose messages align with information operation goals. For example, an opinion article from a site called Climate Change News claimed that the fate of the Amazon rainforest lied in the outcome of the Brazil election. The article suggested that if Bolsonaro was elected, it would be the "tipping point" of the Amazon. This type of content targets audiences that typically follow environmental topics and align with friendlier climate policies to further motivate the potential selection of Lula.

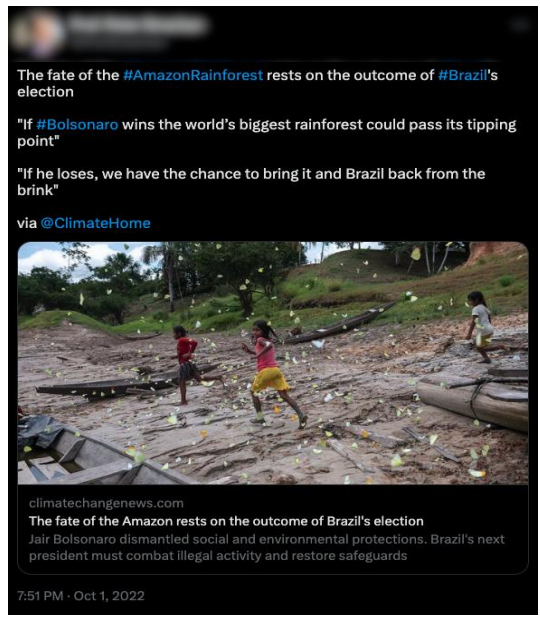

Figure 25: Post showcasing astroturfing technique with an opinion article by Climate Change News weighing the fate of the Amazon Rainforest among the two presidential candidates.

**2.2.4) Conduct Botnet Amplification**

We identified several accounts that seemed to follow well-known bot or cyborg behavior which amplified ideologies, perspectives, or situations aligned with mis- and disinformation narratives. In certain cases, we manually analyzed posting patterns to identify bot-like behavior. For example, one user account shared 73 tweets in one day about Lula. This user account also replied to other posts with spam (e.g., gibberish), memes, and other content demonizing Lula, often repeating the same content in multiple replies without regard to the context. In other cases, we used analysis tools such as TruthNest, BotSentinel, Botometer, FollowerAudit, Heavy.ai to analyze whether accounts were potential bots; TruthNest aggregates over large numbers of tweets to identify common bot behaviors such as spamming, trolling, and prolonged periods of inactivity. For example, one bot user shared the same Bolsonaro images and content multiple times on the same day (Figure 26, Figure 27, & Figure 28). Another piece of evidence for the establishment of online entities for influence purposes is questionable account creation dates and dates of cessation of activity. For example, the account mentioned above joined the platform in April 2022, just as the election season started, and posted its last tweet December 21, 2022.

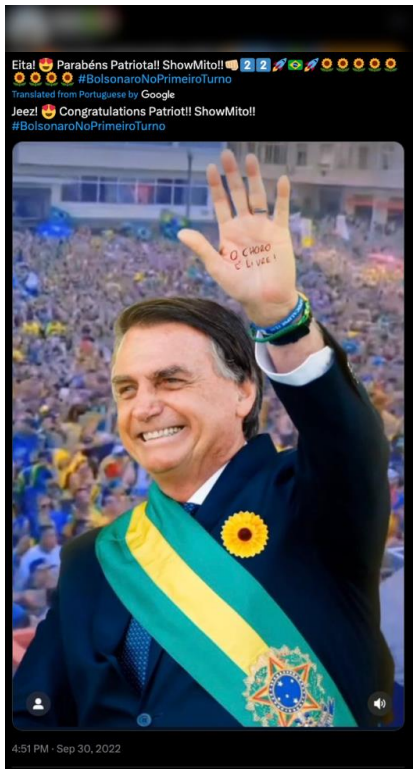
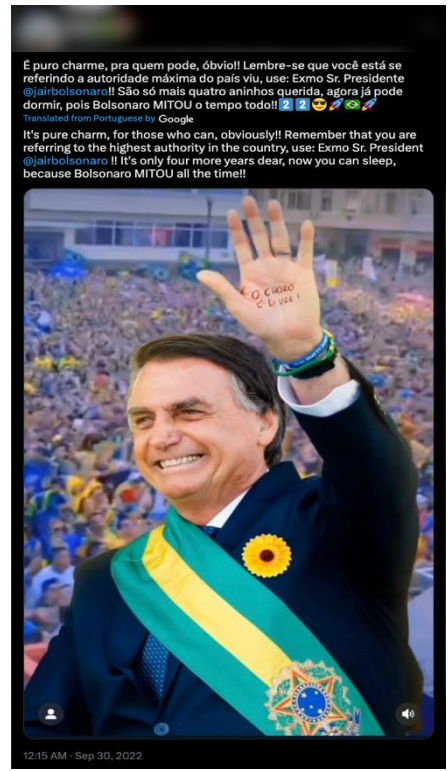

Figure 26: Content likely posted by a bot account showcasing support for Bolsonaro.

Figure 27: Another Tweet published by same account using the same image.

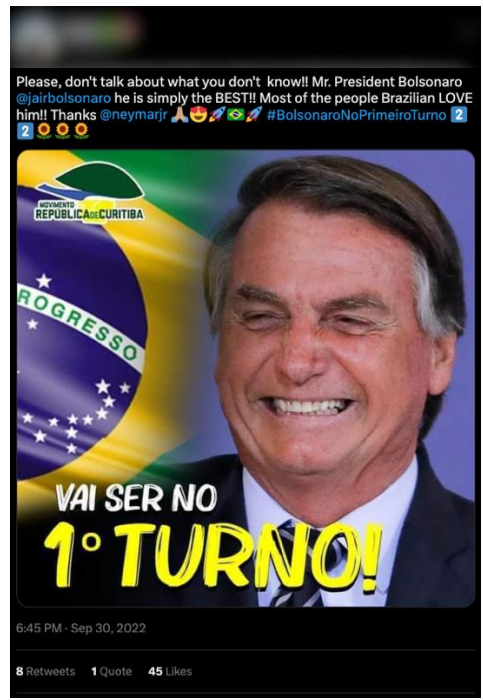

Figure 28: Another Tweet published by same account showcasing further support for Lula. Imagery for the bot accounts are the same and/or similar in nature as evident with this group of figures.

### 2.3) Deliver Content Tactic

### 2.3.1) Post on Platforms Technique

This technique was identified throughout our study, as approximately 97% of our data came from posts on social media sites such as Twitter or Facebook. Additionally, the social media users' ability to create a larger social media footprint within their pages helped prompt their beliefs and narratives. This technique arguably encapsulates one of the easiest yet efficient ways to deliver content to target audience because it allows operatives to exploit the platform features like hashtags to increase views and engagement with their content. For example, one post showcased a meme of Bolsonaro as an Alien, similar in aesthetic to the aliens in the Ridley Scott film, while also taking advantage of hashtags and using a string of them in the post (Figure 29).

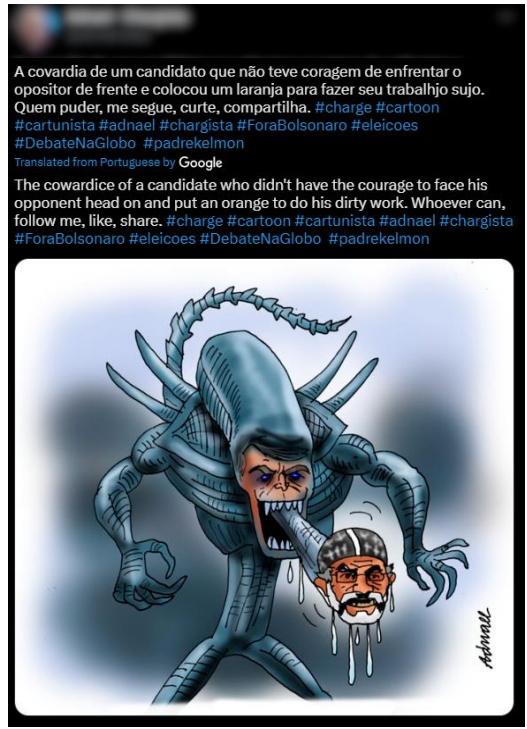

Figure 29: Post with a meme of Bolsonaro appearing as an alien and with accompanying text using hashtags to likely capitalize on how they can be used as social tools.

### 2.3.2) Receive Media Exposure Technique

Media information campaigns are not exclusive to Twitter; state media websites from foreign nations use this strategy attempt to mold public opinion on the election and their candidates (Figure 30). One country whose government-controlled media sites like RT, a Russian state-owned international news network that attributed to approximately ten data points in our research. Given Lula has publicly been sympathetic toward Russia, especially on the topic of the Russo-Ukraine War, it benefits Russian state-controlled media to cover the election and particularly when Lula appeared to be leading in the polls.

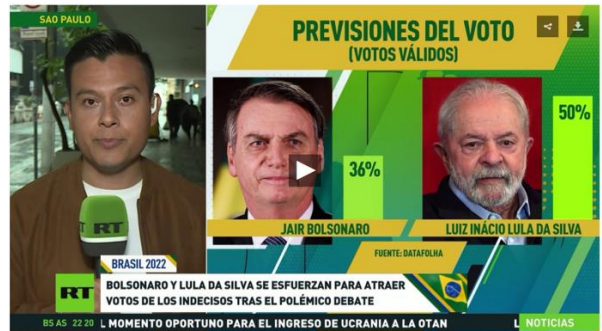

Figure 30: RT en Espanol covering Brazilian election at a point when Lula was leading with votes.

**2.3.3) Leak Documents Technique**

The strategy calls for users posting documentation that misinforms the public on the situation unfolding leading up to the election. The intention of such posts can vary widely. For instance, we observed a post that contained false claims about electoral fraud by the Partido Liberal, Bolsonaro's political party (Figure 31).

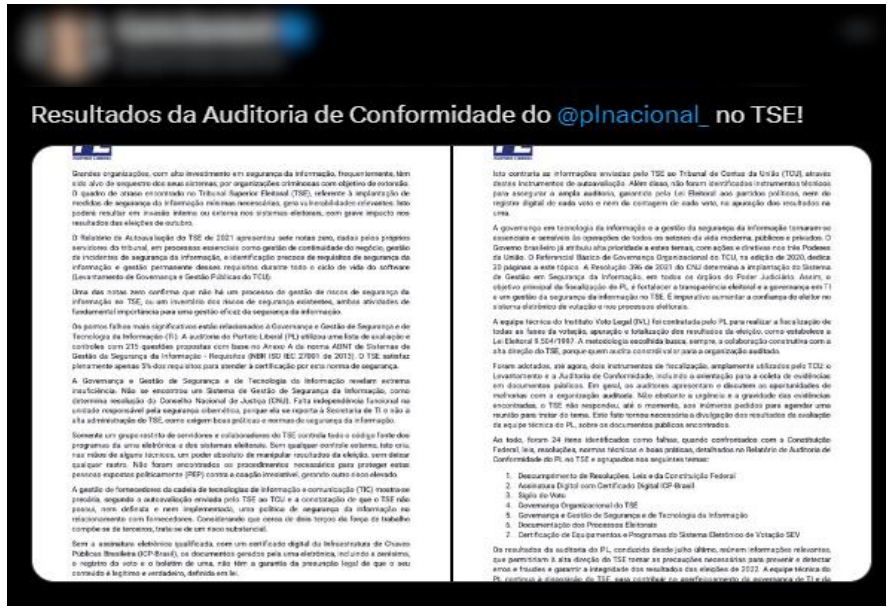

Figure 31: Post with false documents that claim there had been electoral fraud in the Partido Liberal, Bolsonaro's political party.

**2.3.4) Microtargeting Technique**

The strategy called for media accounts to influence support for a certain candidate, political party, or movement by creating a follow-for-follow strategy which increases engagement, followers, and overall perception of these media accounts. For instance, an account may publish a tweet claiming, "Whoever votes for Lula follows me, and I'll follow you back [*sic*]." This behavior aims to create groupings and followings of like-minded individuals and thus contribute to echo chambers of content (Figure 32).

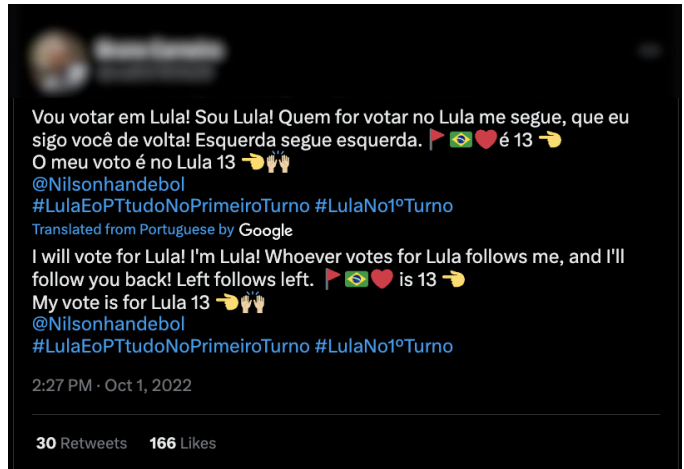

Figure 32: Post showcasing how users aim to create online communities by incentivizing those who are like-minded to follow and like their content.

**2.4) Denigrate Opposing Information Tactic**

**2.4.1) Denigrate Believers of Opposing Narrative Technique**

In this technique, we observed users from both sides of ideological spectrum harass other users. One user posted an article pointing the finger at Neymar for supporting Bolsonaro and saying he has evaded millions in taxes. There were also posts that attempted to villainize supporters of each candidate (Figure 33). This technique capitalized on public shaming to denigrate the opposition. For example, an elected public official, who sides with Lula, shared a video of a male Bolsonaro supporter who humiliated a woman for favoring Lula. Comments to the video showcased backlash to the man's behavior, with some claiming he would be subject to a lawsuit. Others claimed Bolsonaro supporters were trying to blackmail the poor, likely to discourage them from voting.

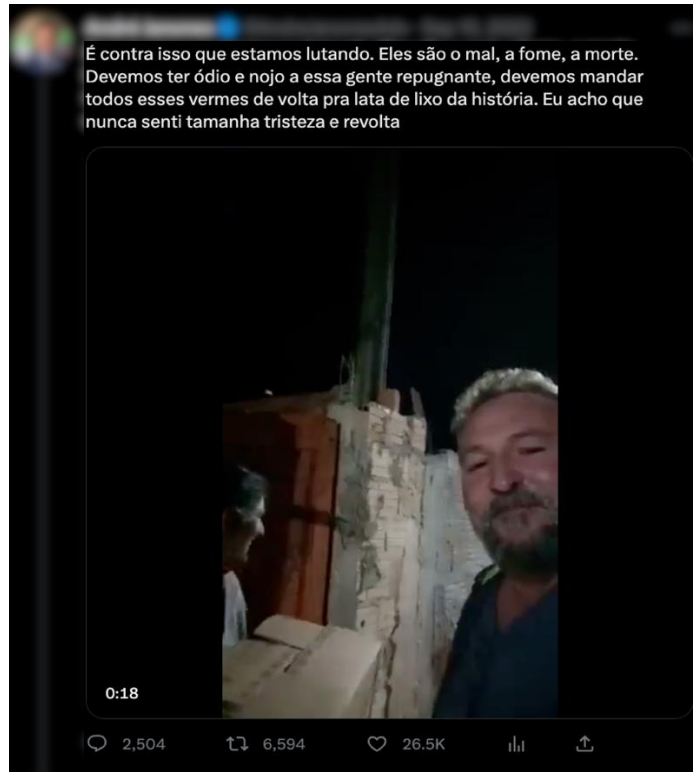

Figure 33: Post showcasing the denigrating the opposition technique in two layers. The first being the account user calling out behavior of Bolsonaro supporter (man) who is recording the video. The second is within the video as the Bolsonaro supporter (man) ridiculed the Lula supporter (woman).

### 2.5) Drive Off-Platform Activity Tactic

### 2.5.1) Drive to Alternative Platforms Technique

In this strategy, we observed accounts on social media sharing links to conduct polls on other platforms. One post encouraged Twitter users to participate in an Instagram poll. The goal with these kinds of posts is to diversify channels for audiences (Figure 34).

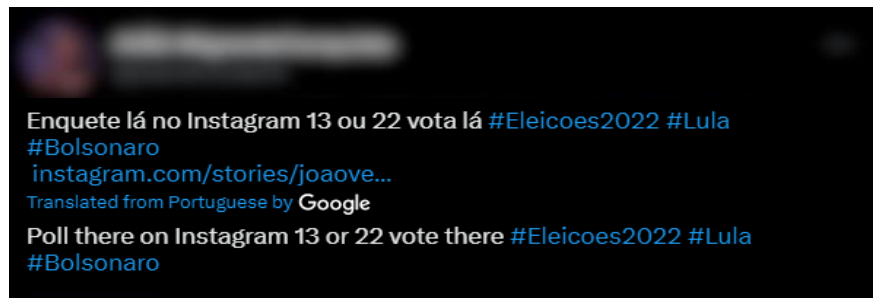

Figure 34: Post showing how some content was meant to drive consumers to other social media platforms.

## Thematic Analysis

We conducted a thematic analysis of the collected posts. This involved categorizing the posts dependent on which candidate it appeared to favor and/or negatively target. We also categorized the content based on the narrative it seemed to contribute to, if relevant. For example, was it catering to narratives about voting machines or a potential rigged election, or other issues such as poverty or the environment.

**Difference in Positive Versus Negative Posts for Bolsonaro and Lula**

Our first observation in the thematic analysis was that there were many more posts in our sample that demonized or ridiculed Bolsonaro (approximately 20% of the sample) than posts doing the same for Lula (only a handful). Anti-Bolsonaro posts predominantly consisted of memes and caricatures. For example, we observed posts where Bolsonaro's face was superimposed onto Hitler's body, and in another post a video placed Bolsonaro side-by-side with a Hitler creating the impression of similarities between the two. Other images showcased Bolsonaro as a baby (to highlight childish behavior) or as a movie villain such as Harry Potter's Voldemort. He was frequently equated to leaders with dictator-like tendencies. The few positive depictions of Bolsonaro put him side by side to Lula and in particular his former arrest. The few posts we found that demonized Lula compared Bolsonaro's army headshot and Lula's former mugshot, communicating a dichotomy between the two candidates equating Bolsonaro as a national hero and public servant while painting Lula as a criminal.

**Narratives Surrounding Legitimacy of Brazil's Electoral System and Voting Machines**

Our next observation was there a common narrative implying the election was rigged or highlighting alleged inefficiencies surrounding Brazil's electronic voting system. Slightly less than half of all posts conveyed these narratives. Mostly, these were posted by people who aligned mostly with Bolsonaro and tried to disseminate the idea the election would be stolen from him and handed to Lula.

**Idolization of Lula**

Finally, Lula was repeatedly idolized by approximately 20% of our artifacts. This support was conveyed in a multitude of ways, such as the users who republished Mark Hamill's post showcasing him as Luke Skywalker. There were also memes of him with a lightsaber destroying Bolsonaro depicted as Darth Vader. In other cases, Lula was superimposed in images with celebrities, or his name was associated with celebrities, to garner attention and sharing on the platform. For example, users edited an image of Taylor Swift to show her with Lula, falsely implying that she supported him. Another post showed Lebron James placing his hands in a L-shape, falsely suggesting that James was supporting or endorsing Lula as a candidate.

**Caveats**

It's important to note this sample size was small and limited by the project budget and timeline. Given this caveat, we cannot make any hard conclusions about the scope of the mis- and disinformation landscape surrounding the Brazilian election. We can say it illustrates how SP!CE coding can be applied to making initial thematic and narrative observations that can help researchers categorize mis- and disinformation, as well as identify trends in the modalities with which they are communicated.

## Best Practices

Before embarking on a data collection and analysis, we recommend extensive studying of the SP!CE framework to familiarize yourself with the tactics and techniques. It is important to note the SP!CE framework covers direct mis- and disinformation as well as other content that can be best described as propaganda or memes, the latter which carry symbolic meaning. The categorization of content can be best observed through a spectrum; due to this, they can sometimes encapsulate multiple techniques based on the perspective and approach of individual researchers. It is important to collaborate amongst team members and supervisors to address any gaps and/or differing perspectives to ensure the content is best catalogued in accordance with the framework.

## Conclusion

In this case study we used the SP!CE 2.1 framework to categorize influence tactics and techniques used by online actors in the 2$^{nd}$ Round of the 2022 Brazilian Presidential election. The categorization of these posts into SP!CE categories within the Enable or Engage phases aided the identification of online users who engaged in mis- and disinformation echo chambers relevant to the two political candidates.

In our research, we observed various patterns of thought, strategy, and tactics employed by malign actors from both sides of the ideological spectrum. For instance, in the subcategory of "Enable," a prevalent technique was the creation of non-synthetic media. Our primary signal for this technique was the use of memes and original visual productions surrounding the election. Memes are especially effective due to their ability to reference pop culture, resonate with online users, and generate a greater probability of user engagement. We observed a trend where most meme artifacts we found supported the Lula campaign and consistently praised Lula while vilifying Bolsonaro.

A second strategy displayed by a significant fraction of the artifacts was the enlistment of intermediaries, specifically the portrayal of celebrities such as Neymar da Silva Santos Júnior, Mark Ruffalo, and Mark Hamill. One of the more interesting developments of this strategy was that significant figures within current-day Hollywood, like Mark Hamill and Mark Ruffalo, actively backed Lula. For example, Hamill produced tweets saying, "Lula 2022," and developed a graphic of Lula as Skywalker from the Star Wars trilogy. Additionally, on the other side of the ideological spectrum we saw Neymar do dances on livestreams supporting the Bolsonaro administration. In its entirety, our research highlighted the significance of pop culture and how these otherwise enigmatic may sway public perception. Nevertheless, these figures were the exception, as other prominent figures such as LeBron James and Taylor Swift had their likenesses used without their apparent consent to gain positive assertions with Pro-Lula or Pro-Bolsonaro accounts.

The most prominent strategy we identified was the distortion of existing narratives through original or existing conspiracy theories. A prevalent theme throughout our collected artifacts was the promulgation of so-called "verifiable" electoral fraud in Brazil. Furthermore, we identified various malign accounts (uniformly in favor of Bolsonaro) spreading these conspiracies via videos, tweets, and memes. In this strategy, we observed a trend of allegedly right-wing accounts endorsing these conspiracies and the usage of false or altered videos and pictures to create the narrative of a corrupted electoral system. While we did not assess the direct impact these narratives had, we could see the alignment in these narratives and those individuals who participated in the truckers' protest and storming the capitol[18,19]. In both instances far-right and Bolsonaro-supporting Brazilians would not accept Bolsonaro's defeat and sought to protest this outcome resulting in blockades and property damages.

In sum, coding specific social media artifacts using the SP!CE framework illustrated the utility of the framework for understanding influence campaigns, and also illustrated how such coding can enable specific thematic and narrative analysis as well as broader general understanding of the trends present in the influence campaign.

## Acknowledgements

This work was funded by MITRE under Master Services Agreement No. 134403, Task Orders 134839, 139839, and 1161636 to Florida International University. We gratefully acknowledge feedback and input from Daniel Sixto and Paul Kim.

---

[18] https://www.pbs.org/newshour/world/brazilian-truckers-protest-bolsonaro-loss-block-hundreds-of-roads
[19] https://www.cfr.org/article/images-show-extent-brazils-capitol-riots